\documentclass[showpacs,preprintnumbers,amsmath,amssymb, nofootinbib]{revtex4}
\usepackage{amsmath}

\begin{document}

\title{Hawking radiation via Anomaly and Tunneling method from Unruh's and Canonical acoustic black hole}
\author{Ram\'{o}n B\'{e}car$^{1}$, Pablo Gonz\'{a}lez$^{2, 3}$, Gustavo Pulgar$^{2}$, and Joel Saavedra$^{2}$}
\address{$^{1}${\small Departamento de F\'{\i}sica, Universidad de Concepci\'on,}
 {\small Casilla 160 C, Concepci\'on, Chile}}
\address{ $^{2}${\small Instituto de F\'{\i}sica, Pontificia
Universidad Cat\'olica de Valpara\'{\i}so, Casilla 4950,
Valpara\'{\i}so}}
\address{ $^{3}${\small Universidad Diego Portales, Casilla 298-V, Santiago, Chile.}}
\date{\today}
\begin{abstract}

We study the Hawking radiation from  Unruh's and Canonical acoustic black
hole from viewpoint of anomaly cancelation method
developed by Robinson and Wilczek and by the simple and physically intuitive picture given by the tunneling mechanism.

\end{abstract}
\maketitle
\section{Introduction}
Hawking radiation is an important quantum effect of black hole
physics. From the quantum point of view, black holes are not
completely black and they can emit radiation with a temperature
given by $h/8 \pi k_{B}GM$ \cite{Hawking:1974sw}, while from a
classical point of view, nothing can escape from the black holes.
Essentially this radiation is thermal and implies black hole could
slowly evaporate emitting quantas. This radiation has been
considered as main tools to understand the quantum nature of
gravity. Hawking radiation shows universality properties and, also
it is determined by universal properties of the event horizon.
There are several approach to obtain the Hawking radiation,  one
of them, was development by Christensen and Fulling
\cite{Christensen:1977jc}, who showed that Hawking radiation in
(1+1)-dimensional Schwarzschild background metric can be derive
from the trace anomaly or conformal anomaly, which arise from the
renormalization of the quantized theory. On recently derivation,
knowing as tunneling method, was proposed in Ref.
\cite{Parikh:1999mf,Srinivasan:1998ty,Shankaranarayanan:2000gb,Shankaranarayanan:2000qv}.
Due its simplicity this method has attracted a lot of attention
and subsequent work, in the tunneling method, Ref.
\cite{Parikh:1999mf,Kraus:1994by,Kraus:1994fj}, one calculated the
imaginary part of the action for the (classically forbidden)
process of s-wave emission across the horizon using the null
geodesic equation \cite{Parikh:1999mf}, which in turn is related
to the Boltzmann factor for the emission at the Hawking
temperature, while the complex path method was first proposed by
K.Srinivasan and T.Padmanabhan \cite{Srinivasan:1998ty} and
subsequently developed by many authors
\cite{Angheben:2005rm,Vagenas:2001qw}. The method developed
employs the Hamilton-Jacobi equations to obtain the particle
classical action along with detailed balance of the ingoing and
outgoing probability amplitudes. The Hamilton-Jacobi method has
been applies to more complicated spacetimes \cite{Sekiwa:2008gk}
and to dynamical black holes \cite{Jiang:1900zz,Yang:2007zze,Di
Criscienzo:2007fm}. However, in Ref.
\cite{Pilling:2007cn,Akhmedov:2006pg}, it was shown that the
WKB/Tunneling method appeared to gave a temperature twice as large
as the correct Hawking temperature. Recently, this problem has
been solved in Ref. \cite{Akhmedova:2008dz,Akhmedov:2008ru}, where
it was shown that in contrast to normal quantum mechanical
WKB/Tunneling problems that the tunneling probability received a
contribution from the time coordinate upon crossing the horizon.
By requiring canonical invariance of the tunneling amplitude and
taking into account the temporal contribution one obtain the
correct Hawking temperature. Another recent approach to the
problem of Hawking radiation was developed by Robinson and Wilczek
in Ref. \cite{Robinson:2005pd}, where the authors combine ideas of
gravitational anomaly and the Hawking radiation of Schwarzschild
type black holes in order to determinate Hawking radiation. This
method has acquired a growing interest and has been applied for
several geometries that described black holes
\cite{Banerjee:2008az,Banerjee:2008ez,Kim:2008hm,
Peng:2008ru,Gangopadhyay:2007hr,Banerjee:2007qs,Setare:2006hq,Wu:2007sw,Miyamoto:2007ue,Chen:2007pp,
Das:2007ru,Kui:2007dy,Iso:2006xj, Vagenas:2006qb, Iso:2006ut,
Iso:2006wa} all those references have showed the robustness of the
anomaly gravitational methods for figure out the nature of
universality,and quantum properties of Hawking radiation. We would
like to note, this effect coming from the event horizon and
therefore is a pure kinematical effect that occurs in any
Lorenzian geometry independent of its dynamical content. This
allow to consider analog models that mimic the properties of black
hole physics. In this sense, the field of analog models of gravity
allows in principle that the most processes of black hole physics
can be studied in a laboratory. In particular the use of
supersonic acoustics flows as an analogy to gravitating systems
was for first the time proposed by Unruh \cite{Unruh:81} and with
the works of Visser \cite{Barcelo:2005fc, Novello:2002qg,
Visser:1998qn, Visser:1997ux, Visser:1993ub} has received an
exponentially growing attention. In this article we computed the
Hawking temperature for two acoustic geometry by the used of the
method developed by Robinson and Wilczek and by the tunneling
method via Hamilton-Jacobi ansatz.

The organization of the paper is as follows: In Sec. II, we
specify the gravitational anomaly method developed by Robinson and
Wilczek. In Sec. III, we discuss as emerged acoustic geometry and
we apply the gravitational anomaly method to Unruh's sonic black
hole. In Sec. IV, we determine the Hawking temperature for the
Canonical acoustic black hole via Robinson-Wilczek method. In Sec.
V, we find the Hawking temperature for both black holes via
tunneling method and finally, we conclude in Sec. VI.

%%%%%%%%%%%%%%%%%%%%%%%%%%%%%%%%%%%%%%%%%%%%%%%%%%%%%%%%%%%%%%%%%%%%%%%%%%%%%%%%%%%%%
%%%%%%%%%%%%%%%%%%%%%%%%%%%%%%%%%%%%%%%%%%%%%%%%%%%%%%%%%%%%%%%%%%%%%%%%%%%%%%%%%%%%%

\section{Robinson-Wilczek method for the Schwarzschild type black hole}
Robinson and Wilczek in Ref.\cite{Robinson:2005pd} have showed that the
energy flux of the Hawking radiation can be fixed by the value of
the gravitational anomaly at the horizon. Here the universality of
Hawking radiation is connected to the universality of the
gravitational anomaly. The starting point was considering a
$d$-dimensional Schwarzschild type space-time with the metric
\begin{equation}\label{metric}
ds^{2}=-f(r)dt^{2}+\frac{1}{f(r)}dr^{2}+r^{2}d\Omega _{(d-2)}^{2},
\end{equation}%
where $d\Omega _{(d-2)}^{2}$ is the line element on the
$(d-2)$-sphere and $f(r)$ is dependent on the matter distribution.
They considered the case where  $f(r)$ has exactly one positive,
real root and all derivatives of $f(r)$ are finite on the
horizon. On the other hand, the contribution to effective action
for the metric $g_{\mu \nu }$ due to matter fields that interact
with this metric is given by
\begin{equation}
W\left[ g_{\mu \nu }\right] \equiv -i\ln \left( \int D\left[
matter\right] e^{iS\left[ matter,g_{\mu \nu }\right] }\right) ,
\end{equation}%
\\
where $S[ matter,g_{\mu \nu }]$ is the classical action functional.
Under general coordinate transformations the classical action $S$ changes as
\begin{equation}\label{condition1}
\delta _{\lambda }S=-\int d^{d}x\sqrt{-g}\lambda ^{\nu }\nabla
_{\mu }T_{\nu }^{\mu }, \
\end{equation}%
where $T_{\nu}^{\mu }$ is the energy-momentum tensor and $\lambda$
is the variational parameter. The symmetry of the classical action (general covariance)
requires that
\begin{equation}
\delta _{\lambda }S=0\Rightarrow \nabla _{\mu }T_{\nu }^{\mu }=0.
\end{equation}%\\
In order to avoid problems of divergence associated
with the Boulware vacuum, Robinson and Wilczek excluded the propagating modes along one
light like direction. The price to pay for this assumption is that the effective
quantum theory is chiral near the horizon and chiral theories
contain gravitational anomalies \cite{Robinson:2005pd}. In consequence, condition (\ref{condition1}) is quantum mechanically violated  due to the chiral
anomaly. Therefore, a real energy-momentum flux it is needed as a compensating object. Main idea of the method,
consist in the divergence of this flux will be canceled the anomaly at the horizon.
 General covariance of the full quantum theory requires
$\delta _{\lambda }W=0$. It is write as
\begin{equation}\label{variationaction}
-\delta _{\lambda }W=\int d^{2}x\sqrt{-g}\lambda ^{\nu }\nabla
_{\mu }T_{\nu }^{\mu }, \
\end{equation}%
where
\begin{equation}\label{emtensor}
T_{\nu }^{\mu }=T_{o\nu }^{\mu }\Theta _{+}+T_{i\nu }^{\mu }\Theta
_{-}+T_{\varkappa \nu }^{\mu }H ,\
\end{equation}%
 $\Theta _{\pm }=\Theta (\pm r\mp r_{H}-\varepsilon )$ are scalar
  step functions and $H=1-\Theta _{+}-\Theta _{-}$.
  The energy-momentum tensors $T_{o\nu }^{\mu }$ and $T_{i\nu }^{\mu} $  are
  covariantly conserved  outside and inside the horizon,
  respectively.  However, $T_{\varkappa \nu }^{\mu }$ in (\ref{emtensor}) is not
  conserved due to the chiral anomaly at the horizon and this anomaly is purely
  time-like and given by \cite{Bertlmann:2000da}
\begin{equation}\label{anomaly}
\nabla _{\mu }T_{\chi \nu }^{\mu }\equiv A_{\nu }\equiv \frac{1}{\sqrt{-g}}%
\partial _{\mu }N_{\nu }^{\mu },
\end{equation}%
where
\begin{equation}
N_{\nu }^{\mu }=\frac{1}{96\pi }\epsilon ^{\beta \mu }\partial
_{\alpha }\Gamma _{\nu \beta }^{\alpha }, \
\end{equation}%
and $\epsilon ^{\beta \mu }$ is the two dimensional Levi-Civita
tensor. Eq.(\ref{variationaction}) can be write as
\begin{eqnarray}\label{variation2}
\nonumber -\delta _{\lambda }W &=&\int d^{2}x\sqrt{-g}\lambda ^{\nu }\nabla _{\mu
}\left\{ T_{\chi \nu }^{\mu }H+T_{o\nu }^{\mu }\Theta _{+}+T_{i\nu }^{\mu
}\Theta _{-}\right\}, \\
\nonumber &=&\int d^{2}x\lambda ^{t}\left\{ \partial _{r}\left( N_{t}^{r}H\right)
+\left( T_{ot}^{r}-T_{\chi t}^{r}+N_{t}^{r}\right) \partial \Theta
_{+}+\left( T_{it}^{r}-T_{\chi t}^{r}+N_{t}^{r}\right) \partial \Theta
_{-}\right\} \\
&&+\int d^{2}x\lambda ^{r}\left\{ \left( T_{or}^{r}-T_{\chi r}^{r}\right)
\partial \Theta _{+}+\left( T_{ir}^{r}-T_{\chi r}^{r}\right) \partial \Theta
_{-}\right\}.
\end{eqnarray}
Form of the $T_{\nu }^{\mu }$ up to an arbitrary function of $r$ and two constants of
integration, $K$ and $Q$, is given by
\begin{eqnarray}
\nonumber T_{t}^{t} &=&-\frac{\left( K+Q\right) }{f}-\frac{B\left( r\right) }{f}-\frac{%
I\left( r\right) }{f}+T_{\alpha }^{\alpha }, \\
T_{r}^{r} &=&\frac{\left( K+Q\right) }{f}+\frac{B\left( r\right) }{f}+\frac{%
I\left( r\right) }{f}, \\
\nonumber T_{t}^{r} &=&-K+C\left( r\right) =-f^{2}T_{r}^{t},
\end{eqnarray}
where
\begin{equation}
B(r)=\int_{r_{H}}^{r} f(x)A_{r}(x)dx, \
\end{equation}%
\begin{equation}
C(r)=\int_{r_{H}}^{r} A_{t}(x)dx, \
\end{equation}%
\begin{equation}
I(r)=\frac{1}{2}\int_{r_{H}}^{r} T_{\alpha }^{\alpha }(x)f^{\prime}
(x)dx.
\end{equation}%
Here, was assumed that
$\frac{I}{f}|_{r_{H}}=\frac{1}{2}T_{\alpha }^{\alpha }|_{r_{H}} $
becomes finite, and
\begin{equation}\label{limit}
\lim_{(r-r_{H})\rightarrow0_{-}}(\frac{1}{f})=-\lim_{(r-r_{H})\rightarrow0_{+}}(\frac{1}{f}).
\end{equation}%
In the limit $ \varepsilon \rightarrow 0 $, using (\ref{limit})
the variation (\ref{variation2}) becomes
\begin{eqnarray}
\nonumber\delta _{\lambda }W &=&\int d^{2}x\lambda ^{t}\left\{ \left[ K_{o}-K_{i}%
\right] \delta \left( r-r_{H}\right) -\varepsilon \left[ K_{o}+K_{i}-2K_{%
\chi }-2N_{t}^{r}\right] \partial \delta \left(
r-r_{H}\right)+...\right\} \\ \nonumber &&-\int d^{2}x\lambda
^{r}\left[ \frac{K_{o}+Q_{o}+K_{i}+Q_{i}-2K_{\chi
}-2Q_{\chi }}{f}\right] \delta \left( r-r_{H}\right)  \\
&&+\int d^{2}x\lambda ^{r}\varepsilon \left[ \frac{K_{o}+Q_{o}-K_{i}-Q_{i}}{f%
}\right] \partial \delta \left( r-r_{H}\right) +... ,
\end{eqnarray}%
where
\begin{eqnarray}
K_{o} &=&K_{i}=K_{\chi }+\Phi , \\
Q_{o} &=&Q_{i}=Q_{\chi }-\Phi ,
\end{eqnarray}%
and
\begin{equation}
\Phi =N_{t}^{r}\mid _{r_{H}},
\end{equation}%
then total energy-momentum tensor reads,
\begin{equation}
\ T_{\nu }^{\mu }=T_{o\nu }^{\mu }\Theta _{+}+T_{i\nu }^{\mu
}\Theta _{-}+T_{\varkappa \nu }^{\mu }H,
\end{equation}%
\\
in the limit $
\varepsilon \rightarrow 0
$, becomes
\begin{equation}
T_{\nu }^{\mu }=T_{c\nu }^{\mu }+T_{\Phi \nu }^{\mu },
\end{equation}%
where $T_{c\nu }^{\mu }$ is the conserved energy-momentum tensor without
 any quantum effects, and $T_{\Phi \nu }^{\mu }$ is a conserved tensor
 with $K=-Q=\Phi$.
Therefore, for a metric of the form (\ref{metric}) the components of $N_{\nu
}^{\mu }$ are
\begin{eqnarray}
N_{t}^{t} &=&N_{r}^{r}=0, \\
N_{t}^{r} &=&\frac{1}{192\pi }\left( f^{\prime 2}+f^{\prime \prime }f\right)
, \\
N_{r}^{t} &=&\frac{-1}{192\pi f^{2}}\left( f^{\prime 2}-f^{\prime \prime
}f\right) .
\end{eqnarray}
so
\begin{equation}\label{fluxSch}
\Phi =N_{t}^{r}=\frac{1}{192\pi }f^{\prime 2}(r_{H}). \
\end{equation}%
In addition, it is well known that the surface gravity $k$ in this case is given by\\
\begin{equation}
k=\frac{1}{2}\frac{\partial f}{\partial r}=\frac{1}{2}f^{\prime
}(r_{H}),
\end{equation}%
which implies the following Hawking temperature
\begin{equation}
T_{H}=\frac{k}{2\pi }=\frac{f^{\prime }(r_{H})}{4\pi }.
\end{equation}%
On the other hand, a beam of massless blackbody radiation moving
in the positive $r$ direction at a temperature $T$ has a flux of
the form
\begin{equation}
\
\Phi =\frac{\pi }{12}T_{H}^{2}.
\end{equation}%
Those results mean the flux required to cancel the gravitational anomaly at the
horizon has a form equivalent to blackbody radiation with a
temperature given by $T=k/(2\pi)$ and this is exactly the Hawking
temperature for this spacetime. Thus, the thermal flux required by
black hole thermodynamics is able of canceling the anomaly.

\section{Unruh's acoustic black hole}
Acoustic geometry or the use of supersonic acoustics flows as an
analogy to gravitating systems was for first time proposed by
Unruh \cite{Unruh:81} and with the Visser's works
\cite{Barcelo:2005fc, Novello:2002qg, Visser:1998qn,
Visser:1997ux, Visser:1993ub} has received an exponentially
growing attention. From the original Unruh's idea considering the
motion of sound waves in the  isentropic and convergent fluid
flow. It is possible to obtain the Eulerian equation of motion
from the action
\begin{equation}
S=-\int d^{4}x\left( \rho \, \dot{\psi} +\frac{1}{2}\rho \,(\nabla
\psi )^{2}+u(\rho )\right),   \label{fluidaction}
\end{equation}
where dot means time derivative, $\rho$ is fluid density, $u$ is the
internal energy density and because we are considering an
irrotational flow $\psi$ represent the velocity potential given by
$\overrightarrow{v}=\nabla \,\psi$. Performing variations of the
action (\ref{fluidaction}) respect to $\psi$ and $\rho$ it is
possible to obtain continuity and Bernoulli equations
respectively. As we mentioned before, the basis of the analogy
between gravitational black hole and sonic black holes comes from
considering the propagation of acoustic disturbances
($\overline{\rho}$ and $\overline{\psi}$) on a barotropic,
inviscid, inhomogeneous and irrotational (at least locally) fluid
flow described by ($\overline{\rho_0}$ and $\overline{\psi}_0$).
 It is well known that the equation of motion for this acoustic
disturbance (described by its velocity potential $\overline{\psi
}$) is identical to the Klein-Gordon equation for a massless
scalar field minimally coupled to gravity in a curved space
\cite{Visser:1998qn, Visser:1997ux, Visser:1993ub, Lepe:2004kv}.
For the disturbance the action (\ref{fluidaction}) up to quadratic
order becomes,
\begin{equation}
S=S_{0}+S_{2,}  \label{s2}
\end{equation}
where $S_2$ described the action of perturbations and it is given
by
\begin{equation}
S_{2}=-\int d^{4}x\left( \frac{1}{2}\rho (\nabla \overline{\psi })^{2}-\frac{%
\rho }{2c^{2}}\left( \overline{\psi }+\overrightarrow{v_0}\cdot
\nabla \overline{\psi }\right) \right).  \label{s22}
\end{equation}
At this point we would like to note that this action can be
written in more elegant form
\begin{equation} S_2=-\int
d^{4}x\sqrt{-g}g^{\mu \nu }\partial _{\mu }\overline{\psi }
\partial _{\nu }\overline{\psi }.\label{s2cs}
\end{equation}
Therefore $S_2$ is completely equivalent to the action for a
massles scalar field (field that described sound perturbation)
propagating in a curved space-time whose acoustic metric $g^{\mu
\nu }$given by
\begin{equation}
g_{\mu \nu }=-\frac{\rho }{c}\left(
\begin{array}{lll}
-1 & \cdots  &\,\,\,\,\,\,-v_{o}^{j} \\
\,\,\,\,\,\vdots  & \ddots  & \,\,\,\,\,\,\,\,\,\,\,\vdots  \\
-v_{o}^{j} & \cdots  & c^{2}\delta ^{ij}-v_{0}^{i}v_{0}^{j}
\end{array}
\right),   \label{metric2}
\end{equation}
where $c$ represent the sound velocity \footnote{Greek indices run
from 0-3, while Roman indices run from 1-3}. In Unruh's work, the
acoustic geometry is described by the following sonic line element
\begin{equation}
ds^{2}=\frac{\rho _{0}}{c}\left( -\left( c^{2}-v_{o}^{r2}\right)
d\tau ^{2}+\frac{c\, dr^{2}}{c^{2}-v_{o}^{r2}}+r^{2}d\Omega
^{2}\right) ,  \label{eq1}
\end{equation}%
where $\rho _{0\text{ }}$ is the density of the fluid, $c$ is the
velocity of sound in the fluid (by simplicity we will assume these
quantities constant) and $v_{0}^{r}$ represent the radial
component of the flow velocity. On the other hand, if then we
assume that at some value of $r=r_{+}$ we have the background
fluid smoothly exceeding the velocity of sound,
\begin{equation}
v_{0}^{r}=-c+a(r-r_{+})+\vartheta (r-r_{+})^{2}, \label{eq2}
\end{equation}%
the above metric assumes just the form it has for a Schwarzschild
metric near the horizon. In this limit our metric
reads as follows
\begin{equation}
ds^{2}=\frac{\rho _{0}}{c}\left( -2ac(r-r_{+})d\tau ^{2}+%
\frac{cdr^{2}}{2ac(r-r_{+})}+r^{2}d\Omega ^{2}\right) ,
\label{eq3}
\end{equation}%
where $a$ is a parameter associated with the velocity of the fluid
defined as $(\nabla\cdot\overrightarrow{v})|_{r=r_{+}}$
\cite{Kim:2004sf}. Note that this geometry was study in Refs.
\cite{Kim:2004sf, Saavedra:2005ug} where the authors studied the
low energy dynamics and obtained the greybody factors for the
sonic horizon from the absorption and the reflection coefficients,
and the quasinormal modes respectively.

In the quantum version of this system we can hope that the
acoustic black hole emits acoustic "Hawking radiation". This
effect coming from the horizon of events is a pure kinematical
effect that occurs in any Lorenzian geometry independent of its
dynamical content \cite{Visser:1997ux}. It is well known that the
acoustic metric does not satisfied the Einstein equations, due to
the fact that the background fluid motion is governed by the
continuity and the Euler equations. As a consequence of this fact,
one should expect that the thermodynamic description of the
acoustic black hole is ill defined. However, this powerful analogy
between black hole physics and acoustic geometry admit to extend
the study of many physical quantities and new method developed for
describe black holes physics, such as Robinson-Wilczek anomaly
gravitational approach,  we are considering in the following the
application of this methods in order to compute the Hawking
temperature for this acoustic geometry. The action ({\ref{s2cs})
for the scalar field $\overline{\psi}$ in the background of
dumb hole described before is given by
\begin{equation}
\nonumber S(\overline{\psi})=\int drd\tau r^{2}\int sen\theta d\theta d\phi \overline{\psi} \left( \frac{1}{%
r^{2}}\partial _{r}(f(r)r^{2}\partial _{r})-\frac{1}{f(r)}\partial
_{\tau }^{2}+\frac{1}{r^{2}}\nabla _{\Omega }^{2}\overline{\psi}
\right), \label{unruh11}
\end{equation}
where $\nabla _{\Omega }^{2}$ represent the Laplacian operator on
unitary two sphere. Passing the radial coordinate to tortoise
coordinate ($r^{\ast }$) whose transformation is defined by
$\frac{\partial r^{\ast }}{\partial r}=\frac{1}{f(r)}$ and
performing the partial waves decomposition $\overline{\psi}
=\sum_{l}\overline{\psi}_{l}Y_{l}\left( \theta ,\phi \right)$,
where $l$ is the collection of angular quantum numbers and
$\overline{\psi}_{l}$ depends on the coordinates $t$ and $r$. One
found that the action near the horizon becomes
\begin{equation}
S\left( \overline{\psi} \right) =\sum_{l}\int d\tau drr^{2}\overline{\psi} _{l}\left[ -\frac{%
1}{f(r)}\partial _{\tau }^{2}+\frac{1}{r^{2}}\partial _{r}\left(
r^{2}f(r)\partial _{r}\right) \right] \overline{\psi} _{l}.
\end{equation}
 Then, physics near the horizon can be described using an infinite
 collection of massless two-dimensional scalar fields in the
 following background
\begin{eqnarray}
\nonumber ds^{2} &=&-f\left( r\right) d\tau ^{2}+\frac{1}{f\left( r\right) }dr^{2}, \\
\phi  &=&r^{2},
\end{eqnarray}
where $\phi$ is the two dimensional dilaton field and the flux is
given by (\ref{fluxSch})
\begin{equation}
\Phi =N_{t}^{r}=\frac{1}{192\pi }f^{\prime 2}(r_{H})=\frac{\tilde{a}%
^{2}}{48\pi } . \label{eq39}\
\end{equation}\
Therefore applying the RW method, we get the following Hawking's
temperature for the Unruh's dumb acoustic black hole
\begin{equation}
 T_{H}=\frac{\tilde{a}}{2\pi}.
\end{equation}
This result coincides with the well-known result of the Hawking
temperature for this geometry \cite{Kim:2004sf} and the thermal
emission is only proportional to the control parameter $a$.

%%%%%%%%%%%%%%%%%%%%%%%%%%%%%%%%%%%%%%%%%%%%%%%%%%%%%%%%%%%%%%%%%%%%%%%%%%%%%%%%%%%%%%%%%%%
%%%%%%%%%%%%%%%%%%%%%%%%%%%%%%%%%%%%%%%%%%%%%%%%%%%%%%%%%%%%%%%%%%%%%%%%%%%%%%%%%%%%%%%%%%%%
\section{Canonical acoustic black hole}
Another solution for the fluid flow is the called Canonical
acoustic black hole. It was introduced by Visser in Ref.
\cite{Visser:1997ux}. In the reference was found a solution with a
spherically symmetric flow of incompressible fluid. This implies
the density $\rho$ is a position independent quantity and the
continuity equation implies that $v \sim 1/r^2$. How was
considered a barotropic equation of state pressure is also an
independent position quantity. Therefore the speed of sound also
is a constant. The metric for  the canonical acoustic black hole
is \cite{Visser:1997ux}
\begin{equation}\label{acousticmetric}\
\
ds^{2}=-c^{2}(1-\frac{r_{0}^{4}}{r^{4}})dt^{2}+(1-\frac{r_{0}^{4}}{r^{4}}%
)^{-1}dr^{2}+r^{2}(d\theta ^{2}+sen^{2}\theta d\phi ^{2}),
\end{equation}
where $r_0=\left(\frac{v\,r^2}{c}\right)^{1/2}$ is a normalization
constant and  without of lack  we can consider $c=1$  and the
metric (\ref{acousticmetric}) can be write as Schwarzschild type,
as follow
\begin{equation}
ds^{2}=-f(r)dt^{2}+\frac{1}{f(r)}dr^{2}+r^{2}d\Omega _{(d-2)}^{2},
\end{equation}
where
\begin{equation}
\ f(r)=(1-\frac{r_{0}^{4}}{r^{4}}),
\end{equation}
physics properties as quasinormal modes of this kind of acoustic
geometry were studied in Refs. \cite{Berti:2004ju} and
\cite{Xi:2007yb}. In order to perform the method of
Robinson-Wilczek we need to write the action ({\ref{s2cs}) for the
scalar field $\overline{\psi}$ in the background of the Canonical
acoustic black hole is
\begin{equation}
S(\overline{\psi}) = \int drdt r^{2}\int sen\theta d\theta d\phi\,\, \overline{\psi} \left( \frac{1}{%
r^{2}}\partial _{r}(f(r)r^{2}\partial _{r})-\frac{1}{f(r)}\partial
_{t }^{2}+\frac{1}{r^{2}}\nabla _{\Omega }^{2}
\right)\overline{\psi},
\end{equation}
where $\nabla _{\Omega }^{2}$ represent the Laplacian operator on
unitary two sphere. Passing the radial coordinate to tortoise
coordinate ($r^{\ast }$) whose transformation is defined by
$\frac{\partial r^{\ast }}{\partial r}=\frac{1}{f(r)}$ and
performing the partial waves decomposition $\overline{\psi}
=\sum_{l}\overline{\psi}_{l}Y_{l}\left( \theta ,\phi \right)$,
where $l$ is the collection of angular quantum numbers and
$\overline{\psi}_{l}$ depends on the coordinates $t$ and $r$. One
found that the action near the horizon becomes
\begin{equation}
S\left( \overline{\psi} \right) =\sum_{l}\int dt drr^{2}\overline{\psi}_{l}\left[ -\frac{%
1}{f(r)}\partial _{t }^{2}+\frac{1}{r^{2}}\partial _{r}\left(
r^{2}f(r)\partial _{r}\right) \right]\overline{\psi}_{l}.
\end{equation}
Therefore physics near the horizon can be described using an
infinite collection of massless two-dimensional scalar fields in
the following background
\begin{equation}
ds^{2} =-f\left( r\right) dt ^{2}+\frac{1}{f\left( r\right)
}dr^{2},
\end{equation}
and the flux is given by (\ref{fluxSch}) \
\begin{eqnarray}
\Phi =N_{t}^{r}=\frac{1}{192\pi }f^{\prime 2}(r_{H})=\frac{1}{12\pi r_{0}^{2}%
}.\end{eqnarray}\ Therefore, we get the following Hawking's
temperature for the canonical acoustic black hole
\begin{equation}
T_{H}=\frac{1}{\pi r_{0}}.
\end{equation}

%%%%%%%%%%%%%%%%%%%%%%%%%%%%%%%%%%%%%%%%%%%%%%%%%%%%%%%%%%%%%%%%%%%%%%%%%%%%%%%%%%%%%%%%%%%
%%%%%%%%%%%%%%%%%%%%%%%%%%%%%%%%%%%%%%%%%%%%%%%%%%%%%%%%%%%%%%%%%%%%%%%%%%%%%%%%%%%%%%%%%%%%
\section{Hamilton-Jacobi Method and Tunneling at Event Horizon}
 We will now review an alternative method for calculating black hole tunneling that makes use of the Hamilton-Jacobi equation as an ansatz \cite{Angheben:2005rm}.
This method ignores the effects of the particle self-gravitation and is based on the work of Padmanabhan and his collaborators \cite{Srinivasan:1998ty,Shankaranarayanan:2000gb,Shankaranarayanan:2000qv}. In general the method involves using the WKB approximation to solve a wave equation. The simplest case to model is scalar particles, which therefore involves applying the WKB approximation to the Klein-Gordon equation. The result, to the lowest order of WKB approximation, is a differential equation that can be solved by plugging in a suitable ansatz. The ansatz is chosen by using the symmetries of the spacetime to assume separability. After plugging in a suitable ansatz, the resulting equation can be solved by integrating along the classical forbidden trajectory, which starts inside the horizon and finishes at the outside observer. Since this trajectory is classically forbidden the equation will have a simple pole located at the horizon. So it is necessary to apply the method of complex path analysis and deflect the path around the pole.
We will apply this method to the acoustics metrics previous.
The Klein-Gordon equation for a scalar field $\Phi$ is:
\begin{equation}
g^{\mu\nu}\partial_{\mu}I\partial_{\nu}\Phi-\frac{m^{2}}{\hbar^{2}}\Phi=0.
\end{equation}
Applying the WKB approximation by assuming an ansatz of the form
\begin{equation}
\Phi\left(t,r,x^{i}\right)=exp\left[\frac{i}{\hbar}I\left(t,r,x^{i}\right)\right],
\end{equation}
and then inserting this back into the Klein Gordon equation will result in the relativistic Hamilton-Jacobi equation to the lowest order in $\hbar$
\begin{equation}\label{eq48}
g^{\mu\nu}\partial_{\mu}I\partial_{\nu}I+ m^{2}=0,
\end{equation}
where m is the mass of the scalar particle.
Considering Eq. (\ref{eq3}), Eq. (\ref{eq48}) can be rewritten as
\begin{equation}\label{eq49}
-\frac{1}{f\left(r\right)}\left(\partial_{t}I\right)^{2}+f\left(r\right)\left(\partial_{r}I\right)^{2}+\frac{1}{r^{2}}\left(\partial_{\theta}I\right)^{2}+\frac{1}{r^{2}\sin^{2}\theta}\left(\partial_{\varphi}I\right)^{2}+m^{2}=0.
\end{equation}
As usual, due to the symmetries of the metric, we can suppose a solution as following form
\begin{equation}
I=-Et+W\left(r\right)+J\left(x^{i}\right),
\end{equation}
therefore we have
\begin{equation}\label{eq51}
\partial_{t}=-E, \partial_{r}I=W'\left(r\right),  \partial_{\theta}I=J_{\theta}, \partial_{\varphi}I=J_{\varphi},
\end{equation}
where $J_{\theta}$ and $J_{\varphi}$ are constant respectively.
Then, putting ({\ref{eq51}) into ({\ref{eq49}), we can get the classical action of an outgoing particle
\begin{equation}
I=-Et+\int\frac{\sqrt{E^{2}-f\left(r\right)\left(\frac{1}{r^{2}}J_{\theta}^{2}+\frac{1}{r^{2}\sin^{2}\theta}J_{\varphi}^{2}+m^{2}\right)}dr}{f\left(r\right)}+J\left(x^{i}\right).
\end{equation}
The metric coefficients for sector 'r-t' alter sign at the two sides of the event horizon. Therefore, the path in which tunneling takes place has an imaginary time coordinate. The ingoing and outgoing probabilities are now given by
\begin{equation}
P_{in}=\left|\Phi_{in}\right|^{2}=exp\left[-\frac{2}{\hbar}\left(-EImt-EIm\int_{C}\frac{dr}{f\left(r\right)}\right)\right],
\end{equation}
and
\begin{equation}
P_{out}=\left|\Phi_{out}\right|^{2}=exp\left[-\frac{2}{\hbar}\left(-EImt+EIm\int_{C}\frac{dr}{f\left(r\right)}\right)\right].
\end{equation}
In the classical limit $(\hbar\rightarrow0)$, we have
\begin{equation}
Imt=-Im\int_{C}\frac{dr}{f\left(r\right)}.
\end{equation}
As a result the outgoing probability for the tunneling particle becomes,
\begin{equation}
P_{out}=\left|\Phi_{out}\right|^{2}=exp\left[-\frac{4}{\hbar}EIm\int_{C}\frac{dr}{f\left(r\right)}\right].\label{eq56}
\end{equation}
The principle of "detailed balance" \cite{Srinivasan:1998ty,Shankaranarayanan:2000gb,Shankaranarayanan:2000qv,Vagenas:2001qw}
for ingoing and outgoing probabilities states that
\begin{equation}
P_{out}=exp\left(-\frac{E}{T_{H}}\right)P_{in}=exp\left(-\frac{E}{T_{H}}\right)\label{57}.
\end{equation}
Comparing eq.({\ref{eq56}) and eq.({\ref{57}) we obtain the temperature for the Unruh's acoustic black hole, which is given by
\begin{equation}
T_{H}=\frac{\hbar}{4}\left(Im\int_{C}\frac{dr}{f\left(r\right)}\right)^{-1}
\end{equation}
and using the expression $f\left(r\right)=2a\left(r-r_{H}\right)$ we obtain
\begin{equation}
T_{H}=\frac{a}{2\pi},
\end{equation}
which is the same result obtained previous, via Robinson-Wilczek method.
Via the same procedure we can to obtain the Hawking temperature for Canonical acoustic black hole coinciding with the previous result (Robinson-Wilczek method).

\section{Conclusions}
In this paper, we have considered a quantum scalar fields in a
Unruh's and Canonical acoustic black hole background. Using two
different methods we computed Hawking temperature of two acoustic
geometries, obtaining the same Hawking temperature. The Robinson
and Wilczek method shown that near the horizon, the physics is
described using an infinite collection of massless 1 + 1
dimensional scalar fields, knowing as phonons for a acoustic
geometry. Also we show the robustness of the anomaly method for
computed the Hawking temperature associated whit the even horizon
of those acoustic geometry. Mainly, we showed that the Hawking
radiation from Unruh's and Canonical acoustic black hole. In case
of  Unruh's black hole we found that the Hawking temperature
coincides with the well-known result of the Hawking temperature
for this geometry \cite{Kim:2004sf}  and the thermal emission is
only proportional to the control parameter $a$. We showed that the
gravitational anomaly that  appears in the Unruh's and Canonical
acoustic black hole background is cancelled by the flux of a 1 + 1
dimensional blackbody   radiation at the Hawking temperature.
Certainly, the only way to probe that our fluxes are really
Hawking fluxes, we need to compare they with the Hawking fluxes
that one obtains from integrating the Planck distribution for a
general charged rotating black hole. In order to avoid
superradiance, we considered fermions in our descriptions. For
fermions, the Planck distribution for blackbody radiation at the
Hawking temperature $T_H$, is given by
\begin{equation}
N_{e,m}(\omega)=\frac{1}{e^{\omega-e\Phi-m\Omega_H}+1}.
\end{equation}
Then, Hawking fluxes can be computed by
\begin{equation}\label{eqlast}
F_M=\int_0^\infty\,\frac{d\omega}{2\pi}\,\omega\,\left(N_{e,m}(\omega)-N_{-e,-m}(\omega)\right),
\end{equation}
where we are including the contribution from the antiparticles.
Then from the evaluation of (\ref{eqlast}) for one uncharged and
static geometric we obtain $F_M=\frac{\pi}{12}T_H^2$, that exactly
match with (\ref{eq39}). Therefore, we can concluded that  the
anomaly method and tunneling method are valid for the acoustic
geometric. We believed that these methods can be useful for give a
thermodynamics description of    acoustics geometries despite that
the dumb black holes do not satisfied Einstein's equation.    We
hope to discus this issue in the near future.

%%%%%%%%%%%%%%%%%%%%%%%%%%%%%%%%%%%%%%%%%%%%%%%%%%%%%%%%%%%%%%%%%%%%%%%%%%%%%%%%%%%%
%%%%%%%%%%%%%%%%%%%%%%%%%%%%%%%%%%%%%%%%%%%%%%%%%%%%%%%%%%%%%%%%%%%%%%%%%%%%%%%%%%%%

\section*{Acknowledgments}
This work was supported by COMISION NACIONAL DE CIENCIAS Y
TECNOLOGIA through FONDECYT \ Grant 11060515 (JS). This work was
also partially supported by PUCV Grant No. 123.789/2007(JS). R. B.
was supported by CONICYT scholarship 2006. P.G. was supported by
Direcci\'on de Estudios Avanzados PUCV. G.P. was partially
supported by MINISTERIO DE EDUCACION through MECESUP Grants FSM
0605

%%%%%%%%%%%%%%%%%%%%%%%%%%%%%%%%%%%%%%%%%%%%%%%%%%%%%%%%%%%%%%%%%%%%%%%%%%%%%%%%%%%%
%%%%%%%%%%%%%%%%%%%%%%%%%%%%%%%%%%%%%%%%%%%%%%%%%%%%%%%%%%%%%%%%%%%%%%%%%%%%%%%%%%%%

\appendix

\end{document}